\documentclass{ptptex}

\usepackage{epsfig}
\usepackage{graphicx}
\usepackage{epsfig,amssymb,amsfonts,verbatim}

\def\nn{\nonumber}
\title{How do we interrogate the electrons without roughing them up?}

\author{%       %Use \scshape  for the family name
Andrew Das \textsc{Arulsamy} and Marco \textsc{Fronzi}%
}

\inst{School of Physics, The University of Sydney, Sydney, New South Wales 2006, Australia}

\abst{Electrons are indistinguishable, but the energy of each electron is different in different materials and if we can exploit this energy, then we can systematically study the changes of electronic properties in non free-electron metals.}

\begin{document}

\maketitle

%\section{Introduction}

\begin{figure}[hbtp!]
\begin{center}
%\begin{figure}[tb!]
%\centering
\scalebox{0.9}{\includegraphics{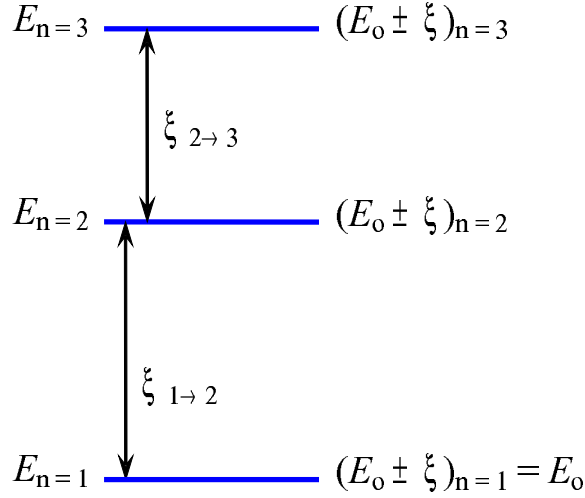}}
%\scalebox{0.45}{\includegraphics{CARL}}
%\includegraphics[width=8mm]{FIG_01}
\caption{Energy levels of hydrogen-like atom (not to scale). $E_{\texttt{n}=1,2,3}$ is the standard energy level notation while, $(E_0 \pm \xi)_{\texttt{n}=1,2,3}$ is the new notation introduced from Eq.~(\ref{eq:1}).}   
\label{fig:1}
\end{center}
%\vspace{0.7cm}
\end{figure}

One of the fundamental requirements from the electronic characterizations of materials is to understand what parameters effectively control the flow of electrons. But we do not have the authority to demand nor persuade the Nature to reveal its secret parameters. Adding to that, all electrons are the same, indistinguishable, we cannot distinguish one from another, be it in a metal or a wood. However, the energy that each or a group of electrons possess is different in different materials due to different types of atoms in a given material. This is also true for electrons in atoms due to different magnitudes of electron-nucleus Coulomb force. In view of this, what we are proposing here is a methodology that, instead of roughing-up the electrons with various experimental techniques, we can interrogate their parent atoms to reveal information about their electronic energies. In other words, the atomic ionization energy or the atomic energy-level difference (see Figure~\ref{fig:1}) can be used to theoretically predict the systematic changes to the conductivity, carrier density, electron-phonon interaction, heat capacity and spin-orbit coupling strength, with respect to different dopant in a given non free-electron compound~\cite{a1,a2,a3,a4,a5}. Such predictions are not only important for experimentalists to evaluate their data and design new materials, but also theoretically significant so as to understand what parameters influence the motion of electrons in strongly correlated matter. The above stated objective (of interrogating the constituent atoms) can be achieved with the methodology proposed in the Refs.~\cite{a1,a2,a3,a4,a5}, which starts with the many-body Hamiltonian,

\begin {eqnarray}
\hat{H}\varphi = (E_0 \pm \xi)\varphi. \label{eq:1}
\end {eqnarray}

$\hat{H}$ is the many-body Hamiltonian, while $\varphi$ is the many-body wavefunction. The eigenvalue, $E_0 \pm \xi$ is the total energy, in which $E_0$ is the total energy at temperature ($T$) equals zero and $\xi$ is the atomic energy-level difference. The + sign of $\pm\xi$
is for the electron ($0 \rightarrow +\infty$) while the $-$ sign
is for the hole ($-\infty \rightarrow 0$). Using this newly defined total energy, we can derive the ionization energy based Fermi-Dirac statistics ($i$FDS), as given below~\cite{a1}

\begin{eqnarray}
&&f_e(E_0,\xi) = \frac{1}{e^{[\left(E_0 + \xi
\right) - E_F^{(0)}]/k_BT }+1}, \nn \\&& f_h(E_0,\xi) = \frac{1}{e^{[E_F^{(0)} - \left(E_0 - \xi
\right)]/k_BT}+1}, \label{eq:2}
\end{eqnarray}

%%%%%%%%%%%%%%%%%%%%%%%%%%%%%%%%%%%%%%%%%%%%%%%%%%%%%%%%%%%%%%%%%%%%%%%%%%%%%%%%%%%%%%%%%%%%%%%%%%%%%%%%%%%%%%%%%%%%%%%%%%%%%
%%%%%%%%%%%%%%%%%%%%%%%%%%%%%%%%%%%%%%%%%%%%%%%%%%%%%%%%%%%%%%%%%%%%%%%%%%%%%%%%%%%%%%%%%%%%%%%%%%%%%%%%%%%%%%%%%%%%%%%%%%%%

%%%%%%%%%%%%%%%%%%%%%%%%%%%%%%%%%%%%%%%%%%%%%%%%%%%%%%%%%%%%%%%%%%%%%%%%%%%%%%%%%%%%%%%%%%%%%%%%%%%%%%%%%%%%%%%%%%%%%%%%%%%%%
\begin{figure}[hbtp!]
\begin{center}
%\begin{figure}[tb!]
%\centering
\scalebox{0.5}{\includegraphics{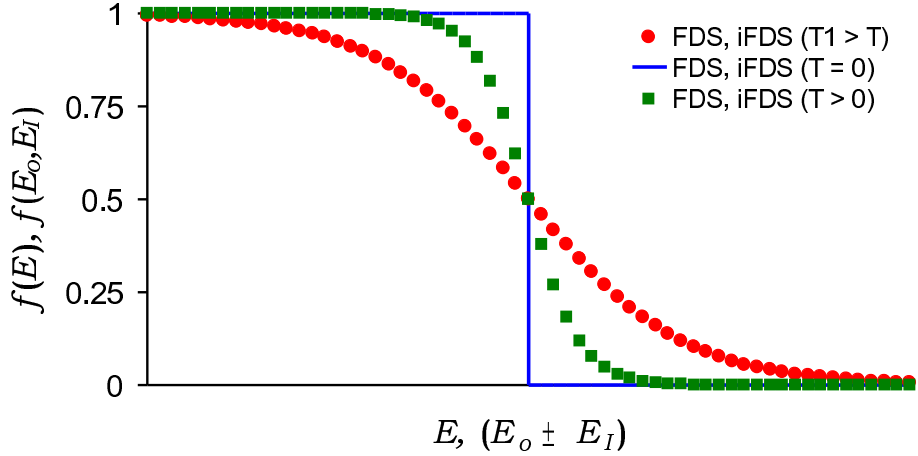}}
\caption{Standard Fermi Dirac (FDS) and ionization energy based Fermi-Dirac ($i$FDS) distributions for temperatures, $T$ = 0, $T >$ 0 and $T1 > T$.}
\label{fig:2}
\end{center}
%\vspace{0.7cm}
\end{figure}
%%%%%%%%%%%%%%%%%%%%%%%%%%%%%%%%%%%%%%%%%%%%%%%%%%%%%%%%%%%%%%%%%%%%%%%%%%%%%%%%%%%%%%%%%%%%%%%%%%%%%%%%%%%%%%%%%%%%%%%%%%%%

where $k_B$ is the Boltzmann constant and $E_F^{(0)}$ is the Fermi level at temperature equals zero. Figure~\ref{fig:2} shows that the standard FDS and $i$FDS are theoretically exact, with non-trivial transformation between them. The explicit content of this operator, $\hat{H}$, namely, the kinetic and potential energy operators need not be known explicitly since $\xi$ is unique for each atom, which can be obtained from the experimental atomic spectra. Using Eqn.~(\ref{eq:2}), one can actually estimate the excitation probability of electrons from different atoms in a given compound. For example, the total energy from Eqn.~(\ref{eq:1}) carries the \textit{fingerprint} of each constituent atom in a compound and it refers to the difference in the energy levels of each atom rather than the absolute values of each energy level in each atom. Hence, the kinetic energy of
each electron from each atom will be captured by the total energy and preserves the atomic level \textit{electronic-fingerprint} in the compound. The application of this theory is well established in strongly correlated matter namely, high-$T_c$ superconductors (cuprates), ferromagnets (manganites), diluted magnetic semiconductors and ferroelectrics (titanates)~\cite{a1,a2,a3}. Physically, $\xi$ implies the energy needed to overcome
the many-body potential energy that exists in a particular system. That is, $\xi$ is the energy needed to excite a particular electron to a finite distance, $r$, not $r \rightarrow \infty$. In solids, the magnitude of $\xi$ is exactly what we need to know that reasonably defines the electronic properties. Hence, $\xi = E_I^{\rm{real}}$ $\propto$ $E_I$, where $E_I$ is the ionization energy of a free atom or ion (with $r \rightarrow \infty$), and its average value can be obtained from~\cite{a1}

\begin{eqnarray}
E_I = \sum_i^z \frac{E_{Ii}}{z}. \label{eq:3}
\end{eqnarray}

The subscripts, $i$ = 1, 2,...$z$, where $z$ denotes the number of valence electrons that can be excited or contributes to the electronic properties of a solid. Recently, various experimental techniques have been employed by Manyala et al.~\cite{manyala} to measure the electronic properties, such as electrical conductivity, magnetic susceptibility and specific heat of Mn doped FeSi compound, and yet, it is not possible to pin-point directly the reasons why and how systematic increment of Mn content (from $x$ = 0.01 to 0.025) increased the conductivity (below 5K) and specific heat of Fe$_{1-x}$Mn$_x$Si. Manyala et al. have indirectly explained the conductivity increment with Mn content as due to increment of temperature-independent mobile carriers. The increment of mobile carriers is assumed from the magnetic susceptibility ($\chi$) results, where $\chi(T)$ curve shifts upward with $x$ ($\chi$ $\propto$ $x$). 

Using our ionization energy theory, we can actually explain how and why the Mn content changed the conductivity and specific heat of Fe$_{1-x}$Mn$_x$Si in the regime that satisfies the under-screened Kondo effect. Firstly, the carrier density ($n$) can be calculated from~\cite{a1} 

\begin{eqnarray}
n = \int_0^\infty{f_e(E_0,E_I)N_e(E_0)dE_0}~\propto ~\exp{\bigg[\frac{-E_I}{k_BT}\bigg]}, \label{eq:4}
\end{eqnarray} 

where $k_B$ denotes the Boltzmann constant whereas, $N_e(E_0)$ is the density of states. The specific heat formula is given by~\cite{a1}

\begin {eqnarray}
C_v = \frac{2\pi^2k_B}{5} \bigg[\frac{k_BT}{\hbar c}\bigg]^3 e^{-\frac{3}{2}\lambda(\xi - E_F^0)} ~\propto ~\exp{\bigg[-\frac{3}{2}\lambda E_I\bigg]}, \label{eq:5}
\end {eqnarray}

where $e$ is the electronic charge, $\lambda = (12\pi\epsilon_0/e^2)a_B$, $\hbar$ is Planck constant, $c$ is the sound velocity, $\epsilon_0$ and $a_B$ denote permittivity of free space and Bohr radius, respectively. Hence, all we need to know now is the relationship between $E_I$ and $x$. Since Mn doped FeSi is equivalent to hole doping~\cite{manyala}, then we can surmise that the average valence state of Mn should be less than Fe. Thus, using Eqn.~(\ref{eq:3}) we obtained the respective averaged values for the ionization energies, $E_I$(Mn$^{2+,3+}$ = 1113.1, 1824.8 kJmol$^{-1}$) $<$ $E_I$(Fe$^{3+,4+}$ = 1760.5, 2642.9 kJmol$^{-1}$). Prior to averaging, the ionization energies for the elements, Mn and Fe were taken from Ref.~\cite{web28}. Smaller $E_I$ implies weak electron-phonon coupling that gives rise to easier electron-flow and also large carrier density. However, this scenario is reversed if $E_I$ is large. Consequently, Mn content (with smaller $E_I$) systematically decreases the $E_I$ of the compound and subsequently increases the carrier density. This in turn shifts the $T$-dependent conductivity (below 5K) and specific heat curves upward with respect to Mn doping in Fe$_{1-x}$Mn$_x$Si, due to Eqn.~(\ref{eq:4}) and Eqn.~(\ref{eq:5}), respectively. In principle, we can use this theory to fine-tune the conductivity in the under-screened Kondo regime or the specific heat of Fe$_{1-x}$Mn$_x$Si with elements other than Mn.

\section*{Acknowledgments}

A.D.A. would like to thank the School of Physics, University of Sydney for the USIRS award and Professors Kostya Ostrikov and Martijn deSterke for their encouragements. This work was also supported by Kithriammah Soosay from Condensed Matter Group, Malaysia. 

%\appendix
%\section{First Appendix} %Empty argument \section{} yields `Appendix'. 
%
%\section{Second Appendix}


\begin{thebibliography}{99}
%%%%%%%%%%%%%%%%%%%%%%%%%%%%%%%%%%%%%%%%%%%%%%%%%%%%%%%%%%%%%
% Some macros are available for the bibliography:
%  o for general use
%    \JL : general journals                 \andvol : Vol (Year) Page
%  o for individual journal 
%    \AJ   : Astrophys. J.           \NC         : Nuovo Cim.
%    \ANN  : Ann. of Phys.           \NPA, \NPB  : Nucl. Phys. [A,B]
%    \CMP  : Commun. Math. Phys.     \PLA, \PLB  : Phys. Lett. [A,B]
%    \IJMP : Int. J. Mod. Phys.      \PRA - \PRE : Phys. Rev. [A-E]     
%    \JHEP : J. High Energy Phys.    \PRL        : Phys. Rev. Lett.
%    \JMP  : J. Math. Phys.          \PRP        : Phys. Rep.
%    \JP   : J. of Phys.             \PTP        : Prog. Theor. Phys.     
%    \JPSJ : J. Phys. Soc. Jpn.      \PTPS       : Prog. Theor. Phys. Suppl.
% Usage:
%  \PRD{45,1990,345}          ==> Phys.~Rev.\ D \textbf{45} (1990), 345
%  \JL{Nature,418,2002,123}   ==> Nature \textbf{418} (2002), 123
%  \andvol{123,1995,1020}    ==> \textbf{123} (1995), 1020
%%%%%%%%%%%%%%%%%%%%%%%%%%%%%%%%%%%%%%%%%%%%%%%%%%%%%%%%%%%%%
  
\bibitem{a1} A. D. Arulsamy, Physica C \textbf{356}, 62 (2001); Phys. Lett. A \textbf{300}, 691 (2002); Phys. Lett. A \textbf{334}, 413 (2005); arXiv:physics/0702232v9; arXiv:0807.0745.

\bibitem{a2} A. D. Arulsamy, X. Y. Cui, C. Stampfl, K. Ratnavelu, Phys. Status Solidi B \textbf{246}, 1060 (2009). 

\bibitem{a3} A. D. Arulsamy, M. Fronzi, Physica E \textbf{41}, 74 (2008).

\bibitem{a4} A. D. Arulsamy, K. Ostrikov, Phys. Lett. A \textbf{373}, 2267 (2009).

\bibitem{a5} A. D. Arulsamy, A. E. Rider, Q. J. Cheng, S. Xu, K. Ostrikov, J. Appl. Phys. \textbf{105}, 094314 (2009).

\bibitem{manyala} N. Manyala, J. F. DiTusa, G. Aeppli, A. P. Ramirez, Nature \textbf{454}, 976 (2008); Z. Fisk, S. von Molnar, Nature \textbf{454}, 951 (2008).

\bibitem{web28} M. J. Winter $\langle $http://www.webelements.com$\rangle $.

\end{thebibliography}
\end{document}